\documentclass[preprint]{aastex}

\usepackage{array}
\usepackage{isotope}
\usepackage{capt-of}
\usepackage{longtable}
\usepackage{multirow}
\usepackage{lscape}
\usepackage{color}
\usepackage[section] {placeins}
\def\apj{ApJ}%
\def\apjl{ApJ}%
\def\aap{A\&A}%
\def\mnras{MNRAS}%
\def\prc{Phys.~Rev.~C}%
\def\nat{Nature}%
\def\gca{Geochim.~Cosmochim.~Acta}%
\def\nar{New~Astr.~Rev.}%
\def\pasa{Publ.~Astronom.~Soc.~Aus.}%

\include{abbrev}
\graphicspath{{}}

\slugcomment{DRAFT: \today}

\shorttitle{Presolar carbon-rich grains: low C and N isotopic ratios}
\shortauthors{}

\begin{document}

\title{Carbon-rich presolar grains from massive stars: subsolar \isotope[12]{C}/\isotope[13]{C} and \isotope[14]{N}/\isotope[15]{N} ratios
and the mystery of \isotope[15]{N}}
\author{M. Pignatari\altaffilmark{1,2,13},
E.~Zinner\altaffilmark{3},
P.~Hoppe\altaffilmark{4},
C.J.~Jordan\altaffilmark{5,14},
B.K.~Gibson\altaffilmark{5,14},
R.~Trappitsch\altaffilmark{6,13}
F.~Herwig\altaffilmark{7,8,13},
C.~Fryer\altaffilmark{9,13},
R.~Hirschi\altaffilmark{10,11,13,14},
F.~X.~Timmes\altaffilmark{12,8,13}
}

\altaffiltext{1}{Konkoly Observatory, Research Centre for Astronomy and Earth Sciences, Hungarian Academy of Sciences, Konkoly Thege Miklos ut 15-17, H-1121 Budapest, Hungary}
\altaffiltext{2}{Department of Physics, University of Basel, Klingelbergstrasse 82, CH-4056 Basel, Switzerland}
\altaffiltext{3}{Laboratory for Space Sciences and Physics Department, Washington University, St. Louis, Mo 63130, USA}
\altaffiltext{4}{Max Planck Institute for Chemistry, D-55128 Mainz, Germany}
\altaffiltext{5}{E.A. Milne Centre for Astrophysics, Dept of Physics \& Mathematics, University of Hull, HU6 7RX, United Kingdom}
\altaffiltext{6}{Department of the Geophysical Sciences and Chicago Center for Cosmochemistry, Chicago, IL 60637, USA.}
\altaffiltext{7}{Department of Physics \& Astronomy, University of Victoria, Victoria, BC, V8P5C2 Canada.}
\altaffiltext{8}{The Joint Institute for Nuclear Astrophysics, Notre Dame, IN 46556, USA}
\altaffiltext{9}{Computational Physics and Methods (CCS-2), LANL, Los Alamos, NM, 87545, USA.}
\altaffiltext{10}{Keele University, Keele, Staffordshire ST5 5BG, United Kingdom.}
\altaffiltext{11}{Institute for the Physics and Mathematics of the Universe (WPI), University of Tokyo, 5-1-5 Kashiwanoha, Kashiwa 277-8583, Japan}
\altaffiltext{12}{Arizona State University (ASU), PO Box 871404, Tempe, AZ, 85287-1404, USA.}
\altaffiltext{13}{NuGrid collaboration, \url{http://www.nugridstars.org}}
\altaffiltext{14}{BRIDGCE UK Network, \url{http://www.astro.keele.ac.uk/bridgce}}

\newcommand{\X}[1]{\textbf{\color{red}{#1}}}

\begin{abstract}
Carbon-rich grains with isotopic anomalies compared to the Sun are found in primi­tive meteorites. 
They were made by stars, and carry the original stellar nucleosynthesis signature. Silicon carbide grains of Type X and C, and low-density graphites condensed in the ejecta of core-collapse supernovae. 
We present a new set of models for the explosive He shell and compare them with the grains showing \isotope[12]{C}/\isotope[13]{C} and \isotope[14]{N}/\isotope[15]{N} ratios lower than solar.
In the stellar progenitor H was ingested into the He shell and not fully destroyed before the explosion. Different explosion energies and H concentrations are considered. 
If the SN shock hits the He-shell region with 
some H still present, the models can reproduce the C and N isotopic signatures in C-rich grains. Hot-CNO cycle isotopic signatures are obtained, including a large production of \isotope[13]{C} and \isotope[15]{N}. The short-lived radionuclides \isotope[22]{Na} and \isotope[26]{Al} are increased by orders of magnitude. The production of radiogenic \isotope[22]{Ne} from the decay of \isotope[22]{Na} in the He shell might solve the puzzle of the Ne-E(L) component in low-density graphite grains. This scenario is attractive for the SiC grains of type AB with \isotope[14]{N}/\isotope[15]{N} ratios lower than solar, and provides an alternative solution for SiC grains originally classified as nova grains. 
Finally, this process may contribute to the production of \isotope[14]{N} and \isotope[15]{N} in the Galaxy, helping to produce the \isotope[14]{N}/\isotope[15]{N} ratio in the solar system. 
\end{abstract}



\section{Introduction}
\label{sec:intro}


Primitive meteorites are carriers of several types of dust of presolar origin, coming from different stellar sources. These sources are identified through measurements of the isotopic abundances in single grains \citep[][]{zinner:14}. 
Presolar C-rich grains that condensed in Core-Collapse Supernovae (CCSNe), 
are low-density (LD) graphite grains \citep[carrier of the Ne-E(L) component,][]{amari:90}, nano-diamonds \citep[carrier of the Xe-HL component,][]{lewis:87}, and silicon carbides (SiC) of Type X \citep[about 1\% of all presolar SiC grains, e.g.,][]{besmehn:03} and of Type C  \citep[0.1-0.2\% of all presolar SiC grains, e.g.,][]{pignatari:13b}. 
Evidence for initial \isotope[44]{Ti} (a prominent signature of SN grains) led to the conclusion that at least one of the few SiC grains originally classified as nova grains is made in CCSNe \citep[][]{nittler:05}. The SiC grains of Type AB \citep[about 4-5\% of all presolar SiC grains,][]{amari:01} possibly have multiple astrophysical sources, including CCSNe \citep[e.g.,][]{zinner:14}.

\isotope[15]{N}-excesses are a signature of SiC grains from CCSNe. They are also found together with \isotope[18]{O} excesses as hotspots, probably linked to TiC subgrains, in LD graphite grains \citep[][]{groopman:12}. The combination of both \isotope[14]{N}/\isotope[15]{N} and \isotope[12]{C}/\isotope[13]{C} ratios lower than solar observed in a fraction of C-rich grains is a challenge to theoretical CCSNe models, assuming mixing between different CCSNe layers and maintaining C/O$>$1 \citep[e.g.,][]{travaglio:99}.
\isotope[15]{N} production in the explosive He-burning shell does not provide a satisfactory solution \citep[e.g.,][]{bojazi:14}.
Presolar SiC grains originally classified as nova grains have \isotope[15]{N}-excesses higher than those of SN grains, in accordance with predictions from nova models. In contrast, a satisfactory explanation for the origin of \isotope[15]{N}-excesses in some SiC-AB grains has not been found yet \citep[][]{huss:97,zinner:14}.

The requirement of a C-rich environment to condense C-rich grains poses a serious problem for finding scenarios that 
explain the observed isotopic abundances in SN grains. \cite{yoshida:07} reproduced the C and N isotopic ratios of a large number of SiC-X grains by mixing matter from different SN layers. Although these mixtures have C/O$>$1 in many cases, the C$>$O constraint and the required heterogeneous and deep mixing in SN ejecta is problematic. To overcome this issue, \cite{clayton:99} proposed formation of C-rich grains from O-rich material. However, \cite{ebel:01} concluded that the O$>$C scenario is unrealistic for SiC formation and \cite{lin:10} showed that the isotope ratios of SiC SN grains are not consistent with predictions for O-rich matter.

In this work we show that the entrainment of H-rich material into the He shell before the SN explosion allows the coproduction of \isotope[13]{C}, \isotope[15]{N} and \isotope[26]{Al}, accounting for isotopic signatures of presolar SiC grains with low \isotope[12]{C}/\isotope[13]{C} ratios.
This also provides a new production scenario for SiC grains classified as nova and AB grains.  

  
The paper is organized as follows. In \S \ref{sec: models_description} we describe stellar models and nucleosynthesis calculations, in \S \ref{sec: comparison} theoretical results are compared with isotope data for different types of presolar grains. In \S \ref{sec:n_galaxy} we discuss the implications of these calculations
for the nitrogen inventory in the Galaxy. In \S \ref{sec: summary} results are summarized.

\section{Stellar model calculations and nucleosynthesis}
\label{sec: models_description}

%

A set of CCSN nucleosynthesis models is calculated from a progenitor $25\mathrm{M_{\odot}}$, $\mathrm{Z}=0.02$
star \citep[][P13 hereafter]{pignatari:13}. 
The simulation of the CCSN explosion includes the fallback prescription by \cite{fryer:12} (model 25d). The initial shock velocity beyond fallback is $\mathrm{v_s}=2\times10^{9}\mathrm{cm}~\mathrm{s}^{-1}$, which corresponds to a total explosion energy of $\mathrm{E_{exp}}=4-7\times10^{51}{\rm erg}$. 
After the end of the central C burning, the convective He shell becomes unstable and eventually H from above the shell is ingested into the He-rich region. The onset of core O burning completely deactivates the convective shell until the SN explosion, leaving He-rich shell material with 
about 1.2\% of H.
The occurrence of mergers between different stellar zones in massive stars is known from the literature \citep[e.g.,][for H mixed into the He shell]{woosley:95}, but its effect on pre- and post-explosive nucleosynthesis has never been investigated before. 
The ingestion of H into hotter layers needs to be simulated with multi-dimensional hydrodynamics because
one-dimensional hydrostatic models lose their predictive power 
\citep[][]{herwig:14}.
In the absence of realistic 3D-hydrodynamics simulations, a qualitative analysis of the effect of residual H during explosive He burning is required.
Together with model 25d from P13 we consider five new models: 25d-H5, 25d-H10, 25d-H20, 25d-H50 and 25d-H500 (models set d). These models share the structure and CCSN explosion of 25d, but the amount of H in the He shell is reduced by a factor of 5, 10, 20 50 and 500 respectively compared to model 25d. 
In P13, the CCSN model with the most extreme conditions is the $15\mathrm{M_{\odot}}$, $\mathrm{Z}=0.02$ model, with $rapid$ SN explosion. The bottom of the C-rich C/Si zone reaches a peak temperature 3.3 times higher than that in the 25d model (2.3$\times$10$^9{\rm K}$ compared to 0.7$\times$10$^9{\rm K}$). The peak density is 100 times larger.
We produced a new set of $25\mathrm{M_{\odot}}$ models, artificially increasing temperature and density to mimic the He-shell conditions of the $15\mathrm{M_{\odot}}$ model, and exploring the same range of pre-supernova H concentrations in He-rich material: 25T-H, 25T-H5, 25T-H10, 25T-H20, 25T-H50 and 25T-H500  
(models set T, see Table\,\ref{table:model_list}).

\begin{figure}
\centering
\resizebox{13cm}{!}{\rotatebox{0}{\includegraphics{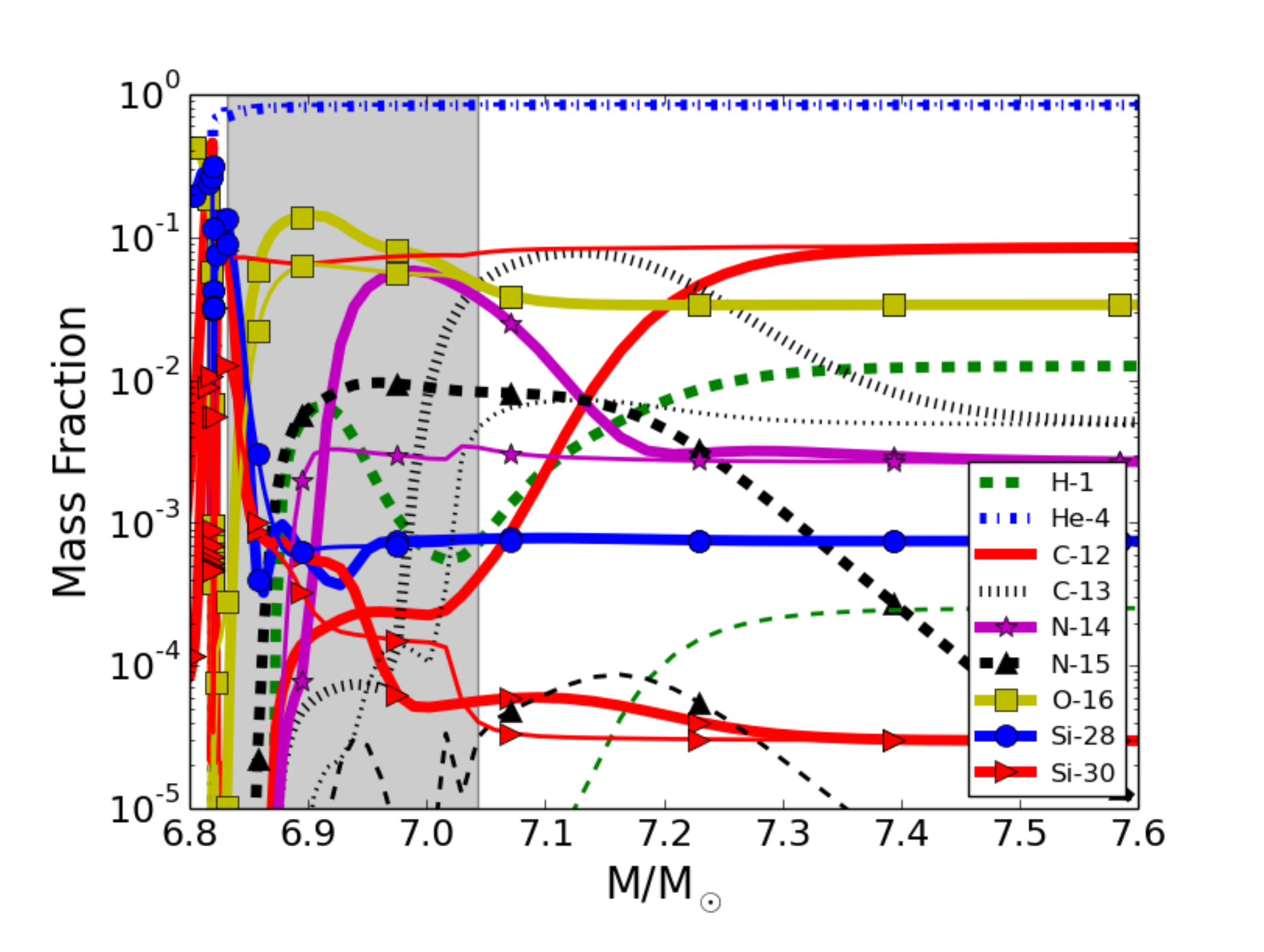}}}
\resizebox{13cm}{!}{\rotatebox{0}{\includegraphics{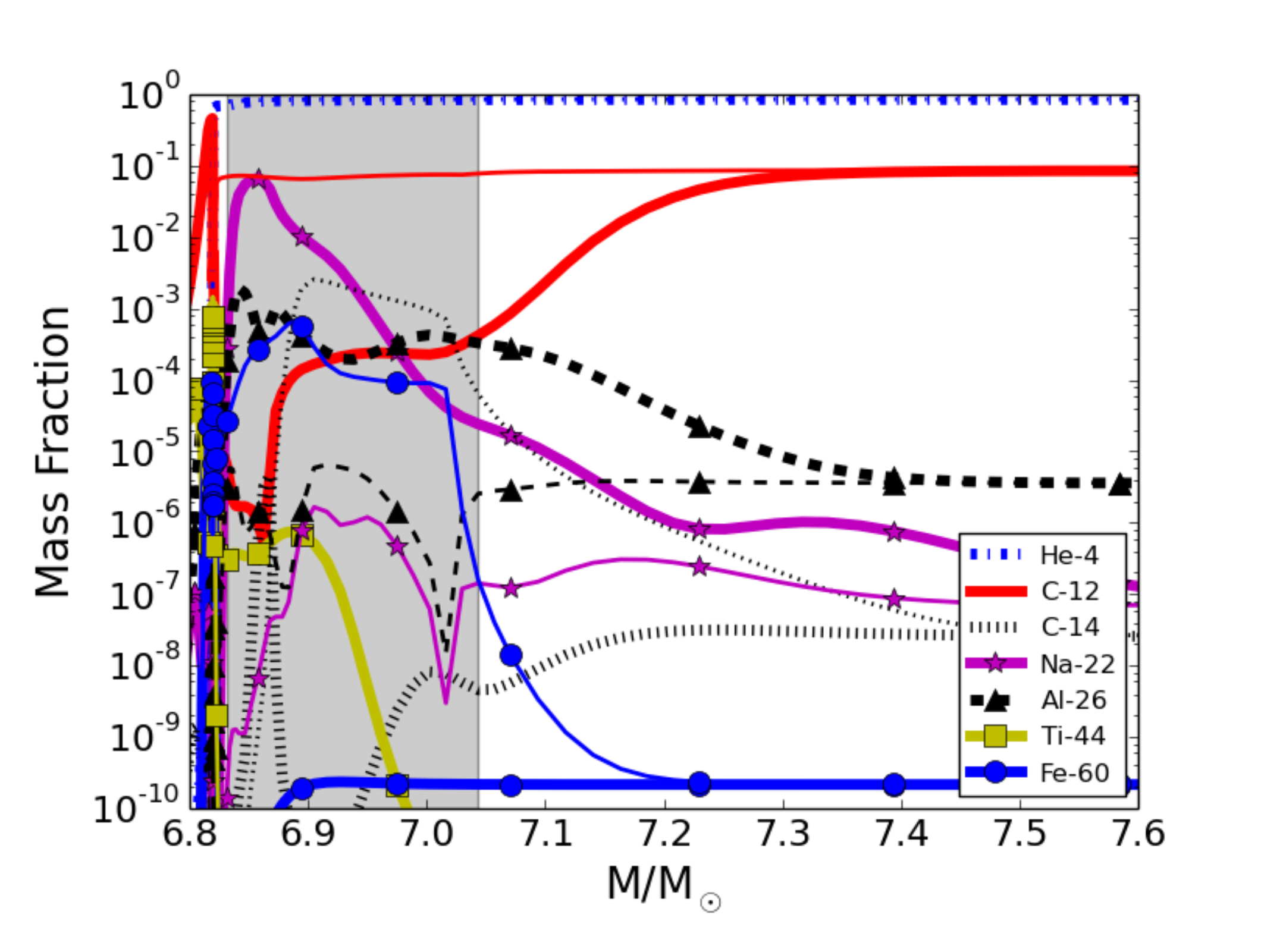}}}
\caption{
Upper panel: Isotopic abundances in the He-shell ejecta of the $25\mathrm{M_{\odot}}$ SN models 25T-H and 25T-H50 (thick and thin lines respectively). Shown are profiles for \isotope[1]{H}, \isotope[4]{He}, \isotope[12,13]{C}, \isotope[14,15]{N}, \isotope[16]{O} and the Si isotopes \isotope[28,30]{Si}. The O-rich O/nova zone that formed in the 25T-H model is highlighted in grey (see text). Markers are used to identify different lines.
Lower panel: we show the short-lived isotopes \isotope[14]{C}, \isotope[22]{Na}, \isotope[26]{Al}, \isotope[44]{Ti} and \isotope[60]{Fe} in the CCSN He-shell ejecta. 
}
\label{fig:abundances_models}
\end{figure}

The post-SN abundances of H, He, C and N isotopes, \isotope[16]{O}, \isotope[28,30]{Si}, and of the short-lived species \isotope[14]{C}, \isotope[22]{Na}, \isotope[26]{Al}, \isotope[44]{Ti} and \isotope[60]{Fe} are reported in Fig.\,\ref{fig:abundances_models} for the models 25T-H and 25T-H50.
Model 25T-H has the largest H concentration and the highest explosion temperature and density. At the bottom of the He shell at $\sim$$6.8\mathrm{M_{\odot}}$, \isotope[28]{Si} is made by $\alpha-$capture on \isotope[16]{O}, forming a C/Si zone \citep[][]{pignatari:13a}. Between 6.83 and $7.04\mathrm{M_{\odot}}$ the abundance of \isotope[12]{C} drops, forming an O-rich zone.
We call this the {\it O/nova zone}. It has C/O$<$1, and the abundances carry an explosive H-burning signature. Qualitatively there are features similar to those of nova nucleosynthesis \citep[e.g.,][]{jose:07}
except in the deepest region of the O/nova zone, which shows unique nucleosynthesis features.
In particular, \isotope[12]{C}(p,$\gamma$)\isotope[13]{N} depletes \isotope[12]{C},
while \isotope[13]{N} feeds \isotope[14]{O} and \isotope[16]{O} via the \isotope[13]{N}(p,$\gamma$)\isotope[14]{O} and the \isotope[13]{N}($\alpha$,p)\isotope[16]{O} reactions. 

Similar nucleosynthesis of explosive H burning is obtained in the 25d models but within a much smaller region ($<$$0.1\mathrm{M_{\odot}}$), and most of the He-shell material is not affected by the explosion. These layers are mostly carrying the pre-SN signature of H burning under He-shell conditions. The \isotope[13]{C} and \isotope[14]{N} abundances are 0.5\% and 0.3\% respectively. A mass fraction of a few 10$^{-6}$ and 10$^{-7}$ is obtained for \isotope[26]{Al} and \isotope[22]{Na}.

In the model 25T-H protons 
suppress the SN neutron burst in the He shell \citep[e.g.,][]{meyer:00}. Protons power the \isotope[22]{Ne}(p,$\gamma$)\isotope[23]{Na} reaction, in competition with the \isotope[22]{Ne}($\alpha$,n)\isotope[25]{Mg} reaction. 
The presence of pre-explosive \isotope[13]{C} provides the \isotope[13]{C}($\alpha$,n)\isotope[16]{O} alternative neutron source, mitigated by the presence of the neutron poison \isotope[14]{N}(n,p)\isotope[14]{C}, and by the competition of the \isotope[13]{C}(p,$\gamma$)\isotope[14]{C} reaction.
The H-burning products \isotope[22]{Na} and \isotope[26]{Al} are boosted (and X(\isotope[44]{Ti})$\lesssim$10$^{-6}$ is made from proton captures on Ca isotopes), while neutron-burst products such as \isotope[60]{Fe} are reduced (Fig.\,\ref{fig:abundances_models}, bottom panel). 
This effect 
depends on the explosion energy and on the H abundance. In model 25T-H the \isotope[26]{Al} peak 
is three orders of magnitude larger than in model 25T-H50. The \isotope[22]{Na} peak is about five orders of magnitude larger. On the other hand, \isotope[60]{Fe} is efficiently produced only in model 25T-H50.
These results might have implications for the observation of the $\gamma-$emission from the decay of \isotope[26]{Al} and \isotope[60]{Fe} in the Galaxy and for their relative abundances 
\citep[][]{diehl:06}. 

In our simulations \isotope[22]{Na} is made in the O/nova zone 
by proton captures, with \isotope[20]{Ne} being made by $\alpha$-capture on \isotope[16]{O} as the main seed. In model 25T-H the abundance peak for \isotope[22]{Na} is almost 10\%.
%
This scenario may explain the Ne-E(L) component in LD graphite grains \citep[][]{amari:90}. This component consists mostly of \isotope[22]{Ne}, suggesting that it is of radiogenic origin from \isotope[22]{Na} originally condensed into the grains. In baseline massive star models, \isotope[22]{Na} is made by C burning in the deeper O-rich O/Ne zone, however it was unclear how to mix Na into the C-rich zones without admixing large quantities of O.
For the first time, we identified a potential pathway to produce \isotope[22]{Na} in the He-shell layers.

\section{Comparison with presolar grains}
\label{sec: comparison}

In this section, the nucleosynthesis models described in \S \ref{sec: models_description} are compared with measurements of different types of C-rich grains with \isotope[12]{C}/\isotope[13]{C} ratios lower than solar
from the Washington University Presolar Grains Database \citep[][and original references in the database]{hynes:09}.
We focus on the abundances in the C-rich regions in the He shell (the C/Si zone and the C-rich He/C zone above) and in the O/nova zone, which lies in between. 

\begin{figure}
\centering
\resizebox{13.cm}{!}{\rotatebox{0}{\includegraphics{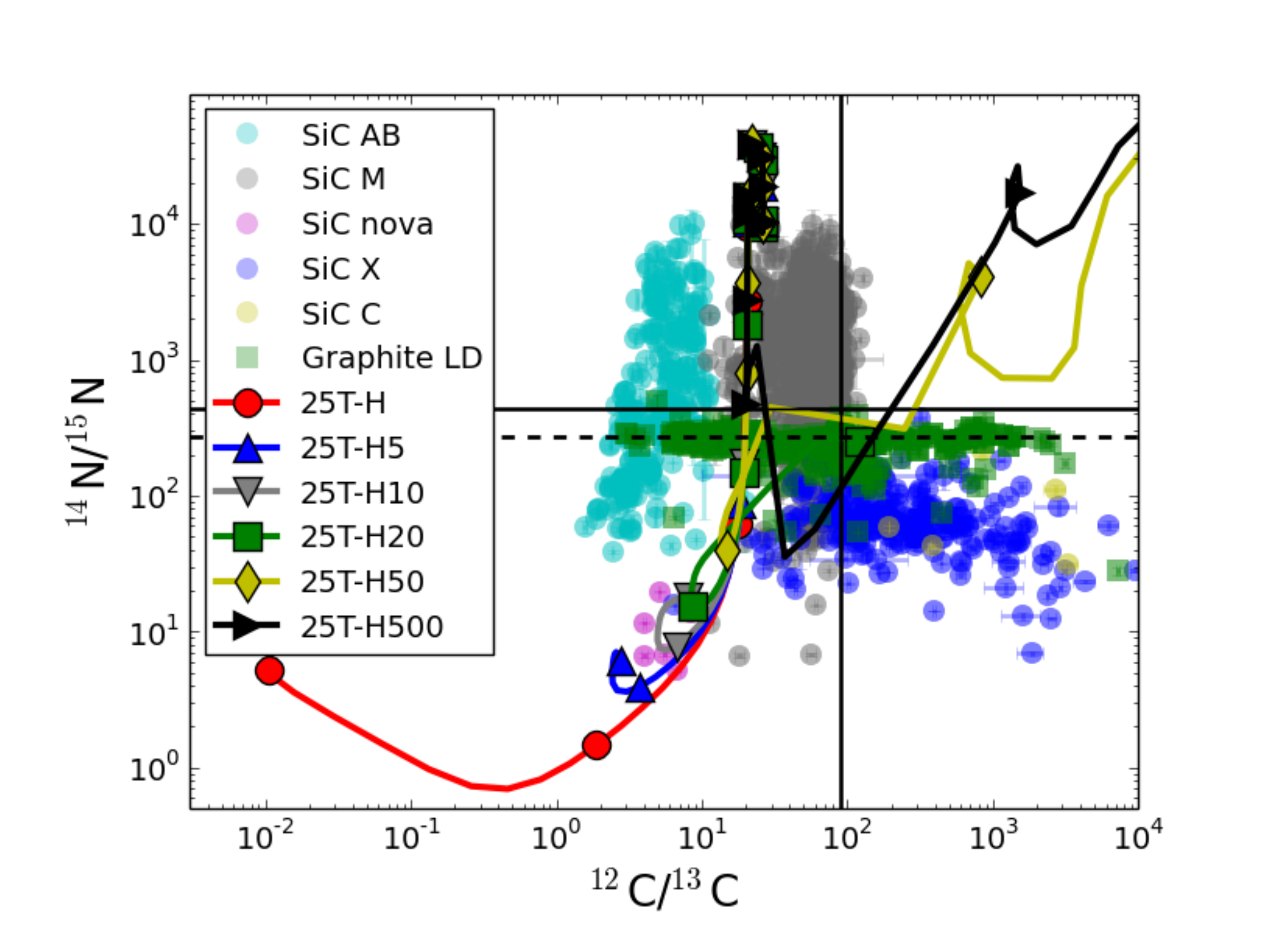}}}
\resizebox{13.cm}{!}{\rotatebox{0}{\includegraphics{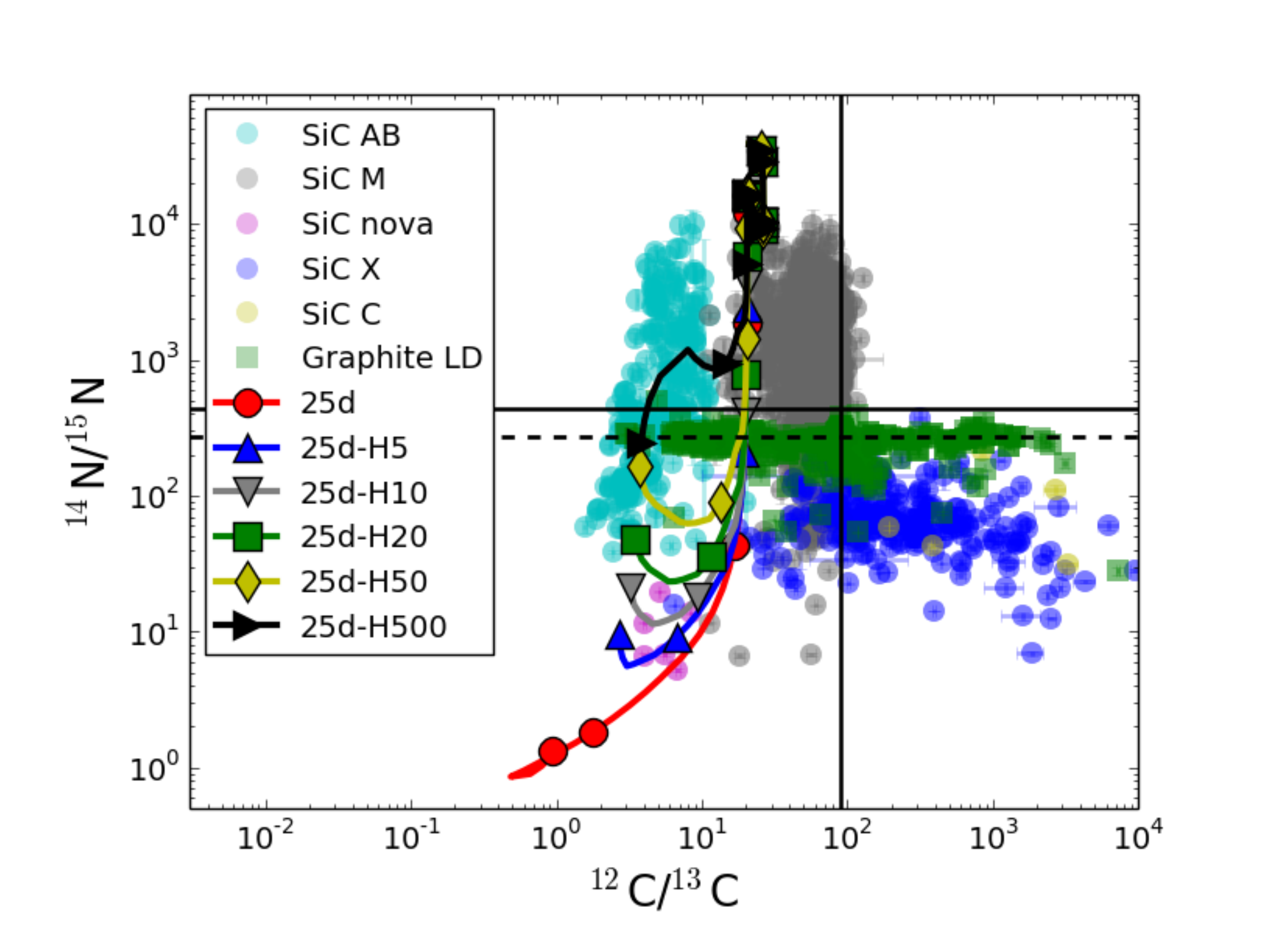}}}
\caption{
Isotopic ratios \isotope[14]{N}/\isotope[15]{N} and \isotope[12]{C}/\isotope[13]{C} across the He/C zone predicted by the sets of models 25T (upper panel) and 25d (lower panel)
are compared with measured data for single SiC grains (type AB, mainstream, nova, X and C) and LD graphites. Errors of grain data are shown (1-sigma).
The solid vertical and horizontal lines indicate the solar isotopic ratios for C and N. For N we used \cite{marty:11}. The terrestrial N ratio (\isotope[14]{N}/\isotope[15]{N}=272) is also shown (dashed horizontal line).
}
\label{fig:iso_grains}
\end{figure}

In Fig.\,\ref{fig:iso_grains}, C and N isotopic ratios of presolar SiC grains are shown together with the predictions from our models in the He shell (He/C zone). 
For \isotope[12]{C}/\isotope[13]{C} lower than solar, the theoretical curves for the models 
25T-H5, 25T-H10, 25T-H20, 25d-H5, 25d-H10, and 25d-H20 reproduce the ratios observed in nova grains, and are compatible with X grains with low N isotopic ratios assuming some degree of dilution with material of close-to-normal composition.
%
%
The 25T-H model produces the lowest \isotope[12]{C}/\isotope[13]{C} ratios, due to large contributions from radiogenic \isotope[13]{C} from \isotope[13]{N} (Fig.\,\ref{fig:abundances_models}).
The 
25T-H5, 25T-H10, 25T-H20, 25d-H5, 25d-H10, and 25d-H20
models show similar \isotope[12]{C}/\isotope[13]{C} ratios. 
The lower limit of the \isotope[14]{N}/\isotope[15]{N} ratio depends on both explosion energy and H concentration. \isotope[15]{N} requires high temperatures to be made as a radiogenic product of \isotope[15]{O}. 
The 25d-H10, 25d-H20, 25d-H50 and 25d-H500 models directly reproduce for the first time the isotopic compositions of SiC-AB grains with N isotopic ratios lower than solar. The 25d and 25d-H5 models show ratios lower than observed, and dilution with normal material is needed. 
Most of graphites show the well-known signature of N isotopic equilibration or contamination \citep[][]{zinner:14}.
Our predictions for C and N ratios might also provide a solution for mainstream SiC grains with \isotope[14]{N}/\isotope[15]{N}$\lesssim$100, not reproduced from low-mass Asymptotic Giant Branch star models \citep[][]{palmerini:11}.
Whether these grains could have condensed from SN ejecta needs to be explored.  

\begin{figure}
\centering
\resizebox{8.1cm}{!}{\rotatebox{0}{\includegraphics{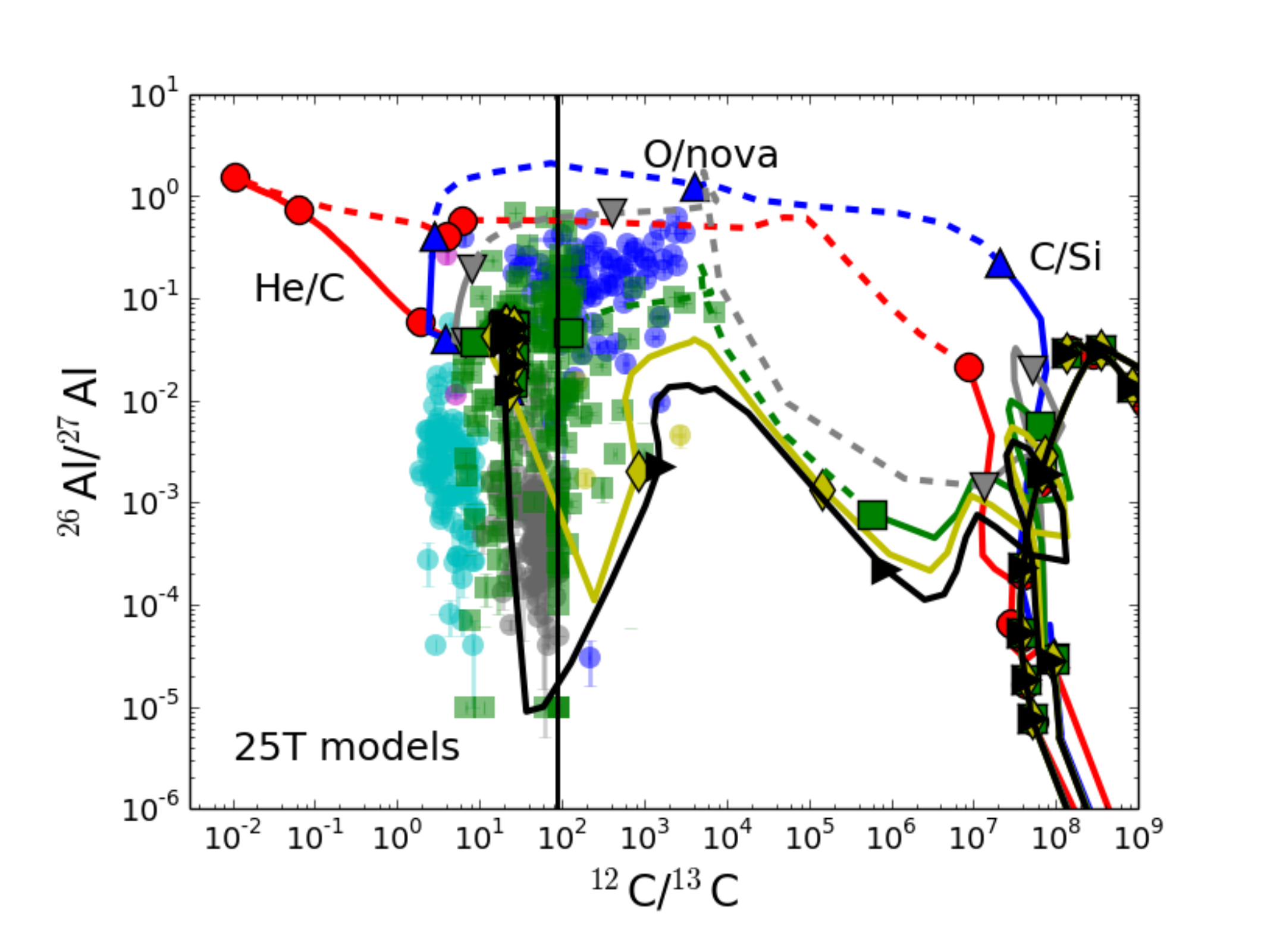}}}
\resizebox{8.1cm}{!}{\rotatebox{0}{\includegraphics{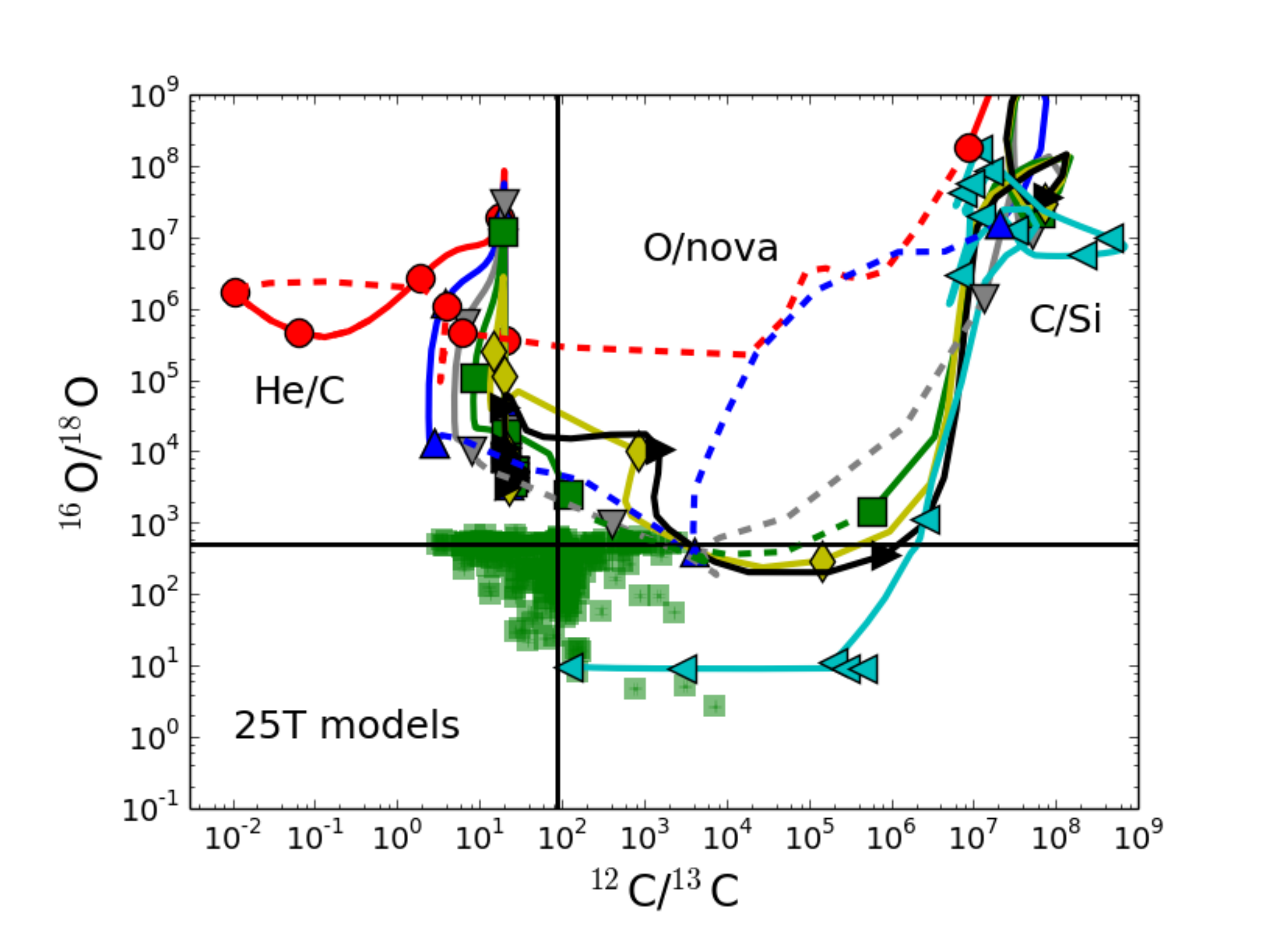}}}
\resizebox{8.1cm}{!}{\rotatebox{0}{\includegraphics{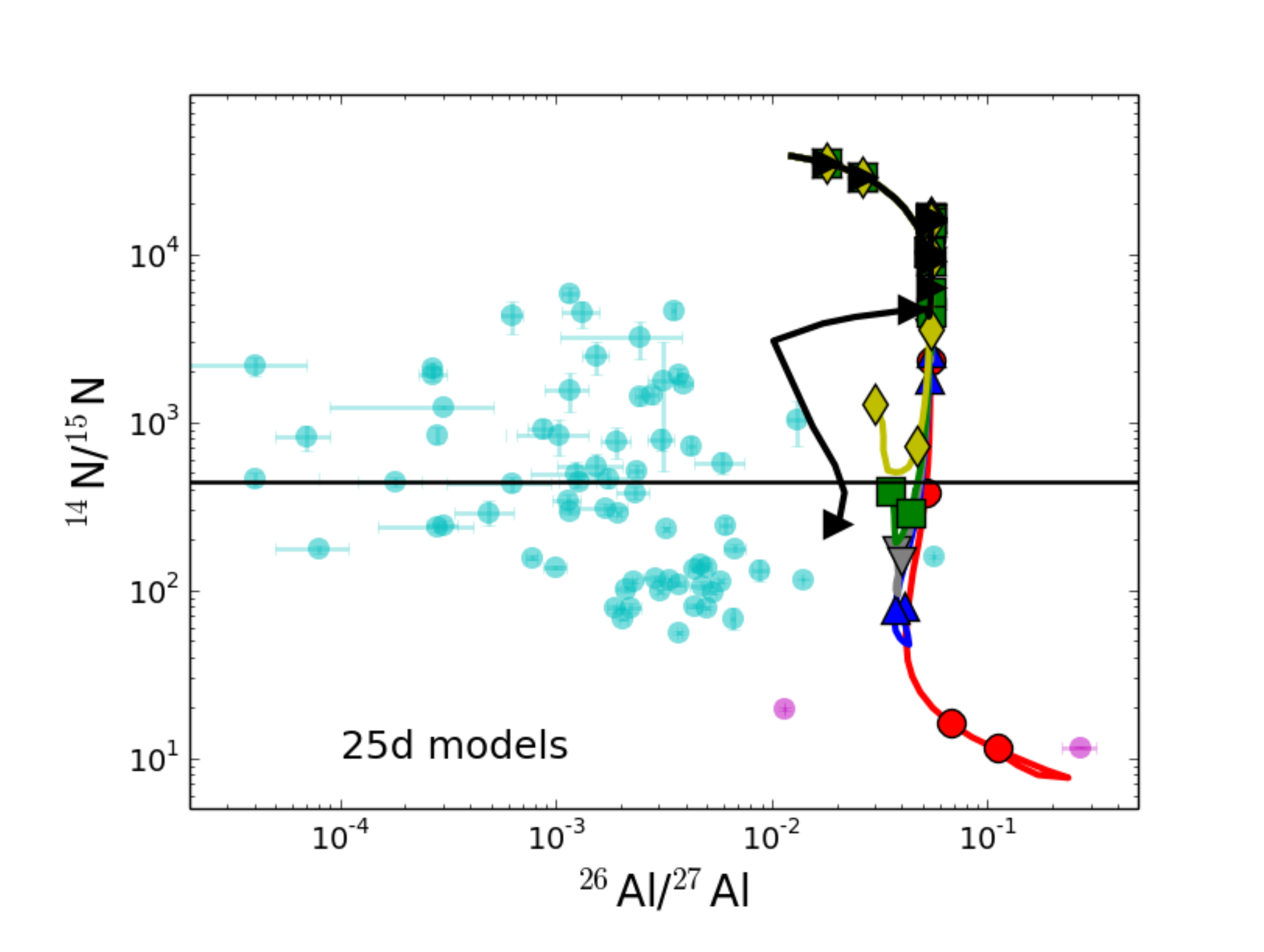}}}
\resizebox{8.1cm}{!}{\rotatebox{0}{\includegraphics{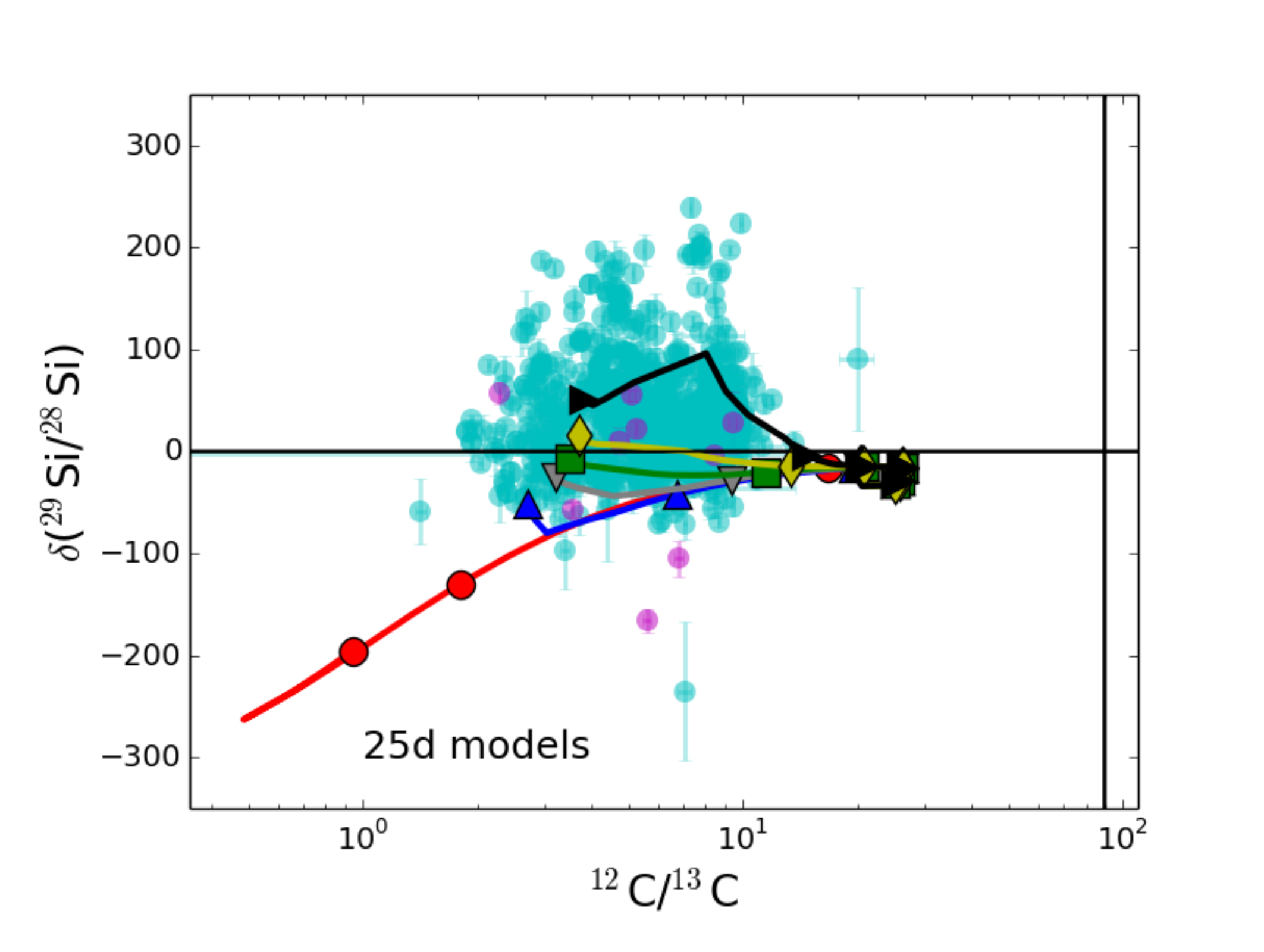}}}
\caption{
Upper Left: Isotopic ratios \isotope[26]{Al}/\isotope[27]{Al} and \isotope[12]{C}/\isotope[13]{C} from the 25T models are compared with those of C-rich grains (see Fig.\,\ref{fig:iso_grains}). Continuous lines represents the C-rich He/C and C/Si zones. The dashed-line part of the curves in between indicates the O/nova zone. 
Upper Right: Isotopic ratios \isotope[18]{O}/\isotope[16]{O} and \isotope[12]{C}/\isotope[13]{C} from the 25T models are compared with LD graphites. The $15\mathrm{M_{\odot}}$ model by \cite{pignatari:13a} is also reported (cyan left-pointing triangles).
Lower Panels: Isotopic ratios \isotope[26]{Al}/\isotope[27]{Al} and \isotope[14]{N}/\isotope[15]{N} (Left Panel) and \isotope[29]{Si}/\isotope[28]{Si} and \isotope[12]{C}/\isotope[13]{C} (Right Panel) from the He/C zone of the 25d models are compared with AB and nova SiC grains. The Si isotopic ratios are in $\delta$ notation ($\delta$(ratio)=(stellar ratio/solar ratio - 1)$\times$1000). For color/symbols of models and grains see Fig.\,\ref{fig:iso_grains}.
}
\label{fig:al26_o18_deltasi29}
\end{figure}

In Fig.\,\ref{fig:al26_o18_deltasi29}, upper-left panel, \isotope[12]{C}/\isotope[13]{C} and \isotope[26]{Al}/\isotope[27]{Al} ratios of C-rich grains are compared with the ratios from the 25T models.
For our models the curves range from high \isotope[12]{C}/\isotope[13]{C} ratios in the C-rich C/Si zone to low \isotope[12]{C}/\isotope[13]{C} ratios in the C-rich He/C zones, with the O/nova zone having \isotope[12]{C}/\isotope[13]{C} ratios in between. 
Therefore, C-rich mixtures from the C/Si, O/nova, and He/C zones of the models 25T-H, 25T-H5, and 25T-H10 can potentially reproduce for the fist time the highest \isotope[26]{Al}/\isotope[27]{Al} ratios observed in SiC-X and LD-graphite grains, consistently with the observed \isotope[14]{N}/\isotope[15]{N} ratio.  
In the upper-right panel we compare model predictions with LD graphites for \isotope[18]{O}/\isotope[16]{O} and \isotope[12]{C}/\isotope[13]{C}. We also report the results for the $15\mathrm{M_{\odot}}$ model \citep[Fig.\,3,][]{pignatari:13a}, which does not include H-ingestion and has CCSN conditions similar to those of the 25T models (\S \ref{sec: models_description}). As discussed in \cite{pignatari:13a}, in the most \isotope[12]{C}-rich layers there is negligible amount of oxygen left and minor mixing with normal material can yield a close-to-normal O isotopic composition.
The effect of the H ingestion is to destroy \isotope[18]{O} and increase \isotope[13]{C}, moving the theoretical curves for the He/C zone from the lower-right quadrant to the upper-right. Unfortunately, a large fraction of graphites show O isotopic equilibration or contamination, in particular for graphites with the lowest \isotope[12]{C}/\isotope[13]{C} \citep[e.g.,][]{zinner:14}, making a meaningful comparison between grain data and models difficult.

The 25d models seem to be more compatible with SiC-AB grains (Fig.\,\ref{fig:iso_grains}).
In Fig.\,\ref{fig:al26_o18_deltasi29}, lower panels, we compare isotopic ratios of AB and nova SiC grains with theoretical curves for 25d models.
AB grains with \isotope[14]{N}/\isotope[15]{N} lower than solar are on average more \isotope[26]{Al}-rich. A nova grain and the AB grains with the highest \isotope[26]{Al} enrichment are compatible with our models, while most of them require some dilution with material of close-to-normal composition. This is consistent with an observed trend of \isotope[26]{Al}/\isotope[27]{Al} inversely proportional to the Al abundance.
Theoretical predictions for \isotope[29]{Si}/\isotope[28]{Si} and \isotope[30]{Si}/\isotope[28]{Si} (not shown in the figure) and C isotopic ratios seem to be consistent with AB and nova grains.

Previously applied SN mixing schemes considered matter from the He/N zone to account for low \isotope[12]{C}/\isotope[13]{C} together with high \isotope[26]{Al}/\isotope[27]{Al} in SN grains \citep[e.g.,][]{zinner:14}. However, in contrast to our 25T and 25d models, those calculations do not provide \isotope[15]{N} in sufficient quantities.
The possibility of having large concentrations of \isotope[13]{C}, \isotope[15]{N} and \isotope[26]{Al} in the He/C zone makes it easier to fit the observed ratios within the constraint of a C-rich mixture in mixing models.

\section{Production of nitrogen in the Galaxy}
\label{sec:n_galaxy}

The peculiar isotopic abundances discussed in the previous sections are relevant for galactic chemical evolution (GCE). While presolar SN grains found in meteorites only condensed from CCSNe that exploded at most several hundred million years before the formation of the Sun \citep[][]{jones:94}, H-ingestion events in the He shell are expected to happen more frequently in low-metallicity stars. 
About 15\% of SiC-X grains and 53\% of LD graphite grains show lower than solar C isotopic ratios (see Presolar grain database).
Therefore, we could argue that at least 15-50\% of the He-shell material ejected from CCSNe was exposed to H ingestion.

We explored the impact of H ingestion on the nitrogen inventory in the Galaxy. Intermediate-mass stars can largely account for the nitrogen abundance in the solar system \citep[e.g.,][]{kobayashi:11}. 
Spectroscopic observations, on the other hand, have shown that massive stars are the most important suppliers of primary nitrogen in the early Galaxy \citep[][]{israelian:04,spite:05}. Recent GCE models underestimate the abundance of \isotope[15]{N} in the Sun by more than a factor of four \citep[][]{kobayashi:11}. Models of fast rotating massive stars can explain the observed production of primary \isotope[14]{N} \citep[][]{chiappini:06}, but not the missing \isotope[15]{N} \citep[][]{meynet:06}. This contrasts with observations of low \isotope[14]{N}/\isotope[15]{N} ratios at high redshift, which suggests a massive star origin for \isotope[15]{N} \citep[$^{14}$N/$^{15}$N=$130\pm20$ at z=0.89,][]{muller:06}, while novae are a controversial source of \isotope[15]{N} at much later timescales \citep[][]{romano:03}.

\begin{figure}
\centering
\resizebox{11.5cm}{!}{\rotatebox{0}{\includegraphics{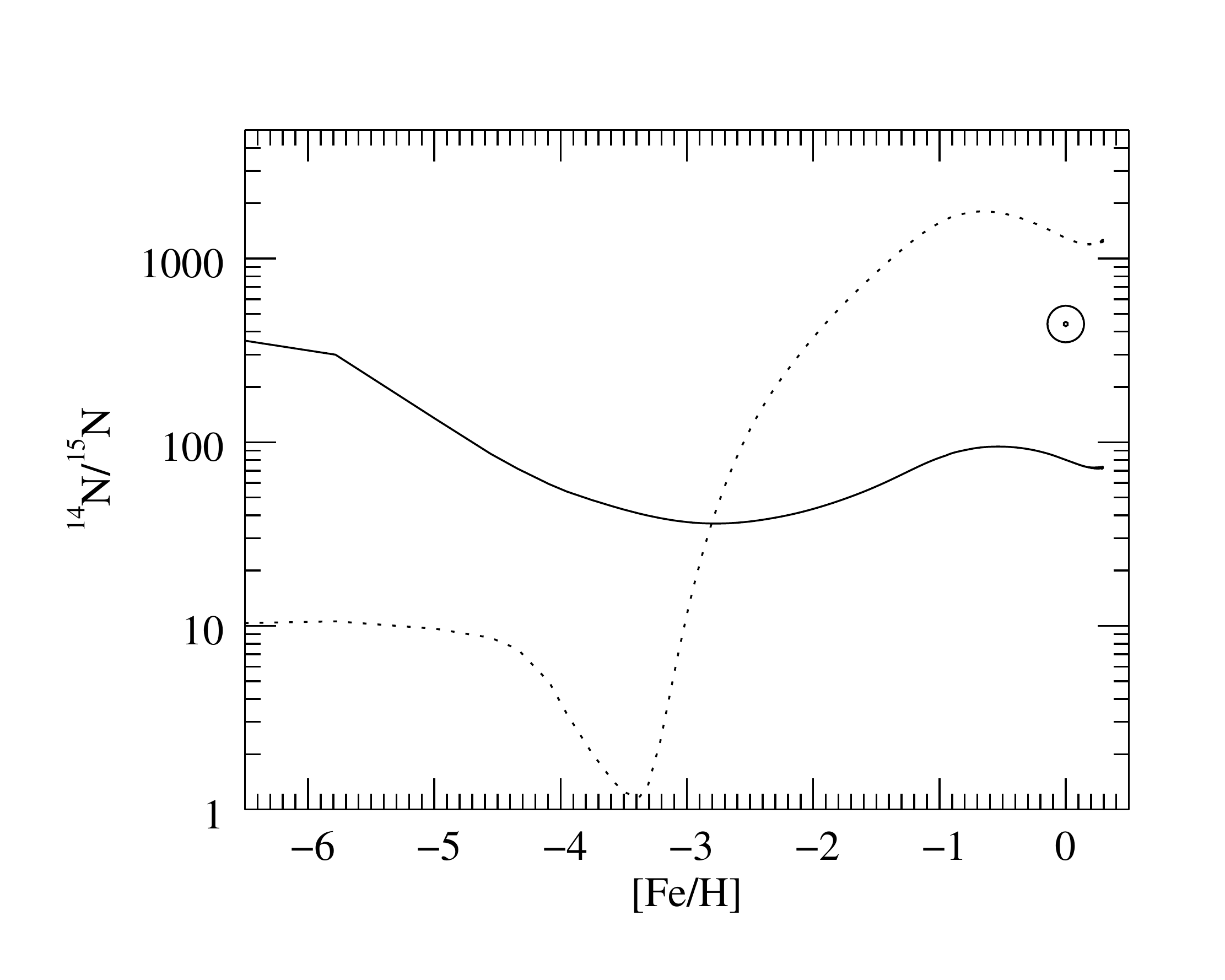}}}
\caption{
GCE simulations of the N isotopic ratio using the CCSNe yields by \cite{woosley:95} (dashed line), and using their solar N yields modified according to the models 25d and 25T-H for all metallicities (continuous line).
For intermediate-mass stars we used the yields by \cite{karakas:10} .
}
\label{fig:n_ratio_gce}
\end{figure}

We tested this scenario using the GCE model presented in \cite{hughes:08}, constructed with the GCE code GEtool \citep[][]{fenner:03}.
We assume that the yields of all CCSNe with initial mass $M<20\mathrm{M_{\odot}}$ carry the N isotopic ratio of model 25T-H (\isotope[14]{N}/\isotope[15]{N}=20), and apply the prediction of model 25d for stars with initial mass $M>20\mathrm{M_{\odot}}$ (\isotope[14]{N}/\isotope[15]{N}=434). Since N is a primary product of H ingestion and is not affected by the initial metallicity of the star, we used the same solar \isotope[14]{N} yields for CCSNe of all metallicities.
The GCE of the N isotopic ratio
is shown in Fig.\,\ref{fig:n_ratio_gce}, in comparison with the baseline case.
With the extreme
assumptions mentioned above, we obtain an N isotopic ratio more than four times {\it lower} than in the Sun. 
Furthermore, our large production of \isotope[14]{N} and \isotope[15]{N} in the He shell of CCSNe changes the GCE results at low metallicities.

We cannot rigorously quantify the impact of stellar yield uncertainties and GCE uncertainties
in Fig.\,\ref{fig:n_ratio_gce} due, in part, to the limitations of current stellar models. Therefore, these results
are meant to be illustrative rather than definitive.
Hydrodynamic simulations of H-ingestion in massive stars are needed to produce more robust stellar yields to quantify the role that these events play in providing a source of nitrogen, and in particular of \isotope[15]{N}, throughout the Galaxy. Major effects can also be expected for other isotopes, above all \isotope[13]{C}.

\section{Conclusions and final remarks}
\label{sec: summary}

We presented new explosive nucleosynthesis calculations for the He shell in CCSNe of massive stars.
The main goal of this work is to study the nucleosythesis impact of the possible presence of H in the He shell, due to H ingestion
before the SN shock reaches these layers.
We explore a large range of explosion conditions (a factor of 3.3 change in the peak temperature of the SN shock, and a factor of 100 in the peak density), and of the H abundance (0.002\%$\lesssim$X$_{\rm H}$$\lesssim$1.2\%) with simplistic 1D SN simulations.


Our calculations have identified the CCSN conditions 
to explain 
puzzling signatures in C-rich presolar grains. Although comparisons of these conditions to the observations can be used to constrain the nature of the supernova explosion, quantitative results require more detailed calculations. 
For example, the multidimensional structure generated by the H-ingestion
cannot be properly represented by 1D-hydrostatic models.
The yield results are sensitive to the peak density and temperature and 3D-hydrodynamics simulations of the H ingestion are needed to quantitatively predict this pre-shock structure. With such detailed models, the features in the C-rich grains can be used to constrain the nature of the CCSN explosion and of SN-shock propagation producing these yields \citep[e.g.][]{wongwathanarat:14}.

We explored the impact of our models on the GCE of the N isotopic ratio, showing that H ingestion in massive stars can be an important source of \isotope[14]{N} and in particular of \isotope[15]{N}, affecting the solar N isotopic ratio. 
To investigate this idea in more detail, massive star simulations with mixing assumptions that allow H-ingestion events
are needed. Such 1D simulations need to be informed by 3D-hydrodynamic simulations, and followed-up with more realistic CCSN simulations. 
For realistic GCE simulations, the yields of 10-12 models of high mass stars at 4 different metallicities (at least) are required. Our results indicate that a similar set of stellar yields from stellar models including H-ingestion events at different metallicities is needed. 


\acknowledgments 
NuGrid acknowledges significant support from NSF grants PHY 02-16783
and PHY 09-22648 (Joint Institute for Nuclear Astrophysics, JINA), NSF grant PHY-1430152 (JINA Center for the Evolution of the Elements) and
EU MIRG-CT-2006-046520. The continued work on codes and in disseminating
data is made possible through funding from STFC and EU-FP7-ERC-2012-St
Grant 306901 (RH, UK), and NSERC
Discovery grant (FH, Canada), and an Ambizione grant of the SNSF
(MP, Switzerland). MP acknowledges support from the "Lendulet-2014" Programme of the Hungarian Academy of Sciences and from SNF (Switzerland).
NuGrid data is served by Canfar/CADC. EZ acknowledges support from NASA grant NNX11AH14G.
BKG acknowledges the support of the UK's Science \& Technology Facilities
Council (ST/J001341/1). RT is supported by NASA Headquarters under the NASA Earth and Planetary Science Fellowship Program through grant NNX12AL85H and was partially supported by the NASA Cosmochemistry Program through grant NNX09AG39G (to A. M. Davis).

\hyphenation{Post-Script Sprin-ger}



\begin{table}
\begin{center}
\caption{List of CCSN models. The He-shell SN ejecta are C-rich, excepting for the O/nova zone in the 25T model set.
Details are given in the text.
}
\begin{tabular}{ccccccc}
\hline
 \multicolumn{1}{m{3cm}|}{{\small X$_{\rm H}$ in the He-shell before the SN shock}} &  1.2\%    &  0.24\%   & 0.12\%    & 0.06\% & 0.024\% & 0.0024\%	 \\
\hline
 \multicolumn{1}{c}{CCSN models set d}   &  25d (P13) & 25d-H5       & 25d-H10   & 25d-H20 & 25d-H50 & 25d-H500     \\
 \multicolumn{1}{c}{He-shell ejecta (M$_{\odot}$)}   & \multicolumn{6}{c}{6.82-9.23}   \\
\hline
 \multicolumn{1}{c}{CCSN models set T}    &  25T-H     & 25T-H5       & 25T-H10   & 25T-H20 & 25T-H50 & 25T-H500     \\
 \multicolumn{1}{c}{He-shell ejecta (M$_{\odot}$)}   & \multicolumn{6}{c}{6.81-9.23}   \\
 \multicolumn{1}{c}{O-rich O/nova (M$_{\odot}$)}   & 6.83-7.04  & 6.84-7.0   &  6.86-7.0  & 6.87-7.0 & - &  -    \\
\hline
 \noalign{\smallskip}
\hline
\end{tabular}
\label{table:model_list}
\end{center}
\end{table}

\end{document}